\documentclass[pdflatex,aps]{revtex4-2}

\usepackage{graphicx}%
\usepackage{amsmath,amssymb,amsfonts}%

\newcommand{\argmax}[2]{\underset{#1}{\mathrm{argmax}}\,#2}

\begin{document}

\title[
Effective interactions behind lift synchronization
]{
Effective attractive and repulsive interactions behind lift synchronization
}

\author{Mitsusuke Tarama}
\email{tarama.mitsusuke@phys.kyushu-u.ac.jp}
\affiliation{Department of Physics, Kyushu University, 
Fukuoka, 819-0395, 
Japan
}

\author{Sakurako Tanida}
\affiliation{Department of Aeronautics and Astronautics, Graduate School of Engineering, The University of Tokyo, 
Tokyo, 113-8656, 
Japan}

\begin{abstract}
Synchronization is a ubiquitous phenomenon in nonequilibrium systems.
One	intriguing example found in every-day life is lifts installed next to each other, that move closely and arrive almost simultaneously during a busy time. 
However, the basic mechanism behind this lift synchronization is yet to be elucidated. 
Here, we investigate the effective interaction acting between the lifts quantitatively. 
Through the analysis on the time-series data obtained by numerically solving a rule-based discrete model of lifts, in which passengers at each floor show up stochastically and call a lift that is expected to arrive first, we find that the effective interaction acting between the lifts consists of not only attraction but also repulsion. 
By changing the parameters of the rule-based model, we are successful to tune the ratio of these competing interactions and to control the dynamics of lifts, realising the transition between in-phase and anti-phase synchronizations. 
Our strategy is applicable to the data of real lifts, and thus it is expected to help controlling lift systems. 
We believe that this study provides a novel approach to design optimal transportation, which is of great importance in improving sustainability of social systems. 
\end{abstract}


\maketitle

\section{Introduction} \label{sec:introduction}
One of the most interesting aspects of active matter is the emergence of various coherent motion in a simple setup~\cite{ Vicsek2012Collective, Elgeti2015Physics, teVrugt2025Metareview}. 
Active matter can interact with each other not only by direct mechanical interactions but also by modifying their surrounding field~\cite{Liebchen2022Interactions}. 
Examples of active matter are found in both animate and inanimate systems, including cytoskeleton~\cite{Sanchez2012Spontaneous,Sekine2024Emergence},  cells~\cite{Tarama2022Mechanochemical}, active colloids~\cite{Bechinger2016Active}, and self-propelled droplets~\cite{Nakata2018Self-organized}. 
Human beings are also active matter~\cite{Gu2025Emergence}. 
Active matter can become \textit{smart} by incorporating information from the surrounding environment~\cite{Doostmohammadi2023Editorial, teVrugt2025Artificial}. 
Although a group of active matter could exhibit coordinated behaviours, the individuals in a group are not necessarily so \textit{smart}; 
That is, individual active matter does not adjust its motion to optimize the collective dynamics, but it rather moves in response to the local condition of its surroundings. 
A recent experiment showed that, although a person is usually smarter than an ant, when working in a group the performance of people to solve a task can be less than that of ants~\cite{Dreyer2025Comparing}. 
Individual biological cells can sense and respond to their complex environment and survive even in dangerous circumstances, while they move collectively, some of them exhibit seemingly inefficient motion such as swirls during wound healing process~\cite{Hakim2017Collective}. 
These are the examples in which individual active matter, being smart when placed alone, fail or cease to exhibit their ability in a group. 

Among the self-organised coherent motion, synchronization is a ubiquitous phenomena in active matter and other non-equilibrium systems that possess oscillatory nature. 
It is first reported by Chrestian Hugens in sixteenth century and has been studied intensively for about this half a century since the seminal studies by Winfree \cite{Winfree1967Biological} and Kuramoto \cite{Kuramoto1975Self-entrainment,Sakaguchi1986A}. 
Examples of synchronization in active matter include rotating flagella and cillia~\cite{Reichert2005Synchronization,Uchida2010Synchronization} and active rotors~\cite{Levis2019Activity,Scholz2021Surfactants,Vahabli2023Emergence}. 
Cardiac cells vibrating spontaneously adjust their phase with each other by mechanical interaction~\cite{Reinhart-King2008Cell-Cell,Nitsan2016Mechanical}, which may help an efficient beating function of a heart. 
It also occurs due to the existence of external confinement~\cite{Sepulchre2005Cooperative,Riedl2023Synchronization}. 
Synchronization has pros and cons, and it may lead to social issues. 
A well-known example is the London millennium bridge, on which people fell in step unintentionally, leading to a huge oscillation of the bridge~\cite{Eckhardt2007Modeling}. 
Therefore, it is of great importance to understand the mechanism behind synchronization not only to avoid unfortunate disasters but also to improve the efficiency. 

One intriguing example that exhibits synchronization 
is lifts installed next to each other~\cite{Poschel1994Synchronization,Hikihara1997Emergent, Tanida2021Dynamic,Tanida2022The, Feng2021When,Nagatani2003Complex}, which one may experience in daily life. 
If there are two lifts, at a busy time they tend to move relatively closely and thus, arrive almost simultaneously. 
The synchronization mechanism of lifts has been intuitively explained in analogy to the clustering or traffic jam of buses~\cite{Helbing2001Traffic,Chowdhury2000Statistical, OLoan1998Jamming,Sugiyama2008Traffic}; 
Suppose there are two buses running relatively close to each other on the same route at a rush hour. 
The forward one has to stop at almost all the bus stops, while the following one has less chance to make a stop because there is not enough time for new passengers to appear after the forward one left a bus stop. 
As a result, the following bus approaches the forward one, resulting in cluster formation. 
However, lifts are different from usual buses not only because they have their own tracks and can pass through each other, but also because they are on-demand system and they move in response to the calls. 
More importantly, the ascent and descent motion of lifts can be highly asymmetric, in particular around the rush hour at the day-start or day-end, when almost all passengers are willing to move from the ground floor to the upper floors or in the opposite direction. 
Although many studies on the dynamics of lifts reported synchronization~\cite{Poschel1994Synchronization,Hikihara1997Emergent, Tanida2021Dynamic,Tanida2022The, Feng2021When} and chaotic motion~\cite{Nagatani2003Complex}, 
the underlying mechanism behind the lift synchronization ---if the same mechanism as the bus-route model is really at play---  is yet to be elucidated. 

The aim of our study is to investigate the mechanism behind lift synchronization. 
To this end, we analyse quantitatively the effective interaction acting between lifts. 
Of course lifts are manufactured system and their motion is controlled by the designed algorithm. 
Some recent ones called \textit{smart elevator systems} are even well designed to optimize their motion by using machine learning. 
However, the lift system that we consider here is not a smart one but a simple one that is ubiquitous in daily life. 
That is, although passengers at each floor call the lift that is expected to arrive first and the lifts react to these demand, the lifts do not communicate directly with each other and they are not controlled to maximize the entire transportation efficiency by a \textit{mother computer}. 
In addition, we focus on the \textit{feierabent} effect~\cite{Poschel1994Synchronization}, namely, the simple situation where all passengers are willing to go to the ground floor, corresponding to the day-end at an office building or the closing time of a department store. 

\section{Results} \label{sec:results}
\begin{figure}[tb]
	\centering
	\includegraphics[width=\textwidth]{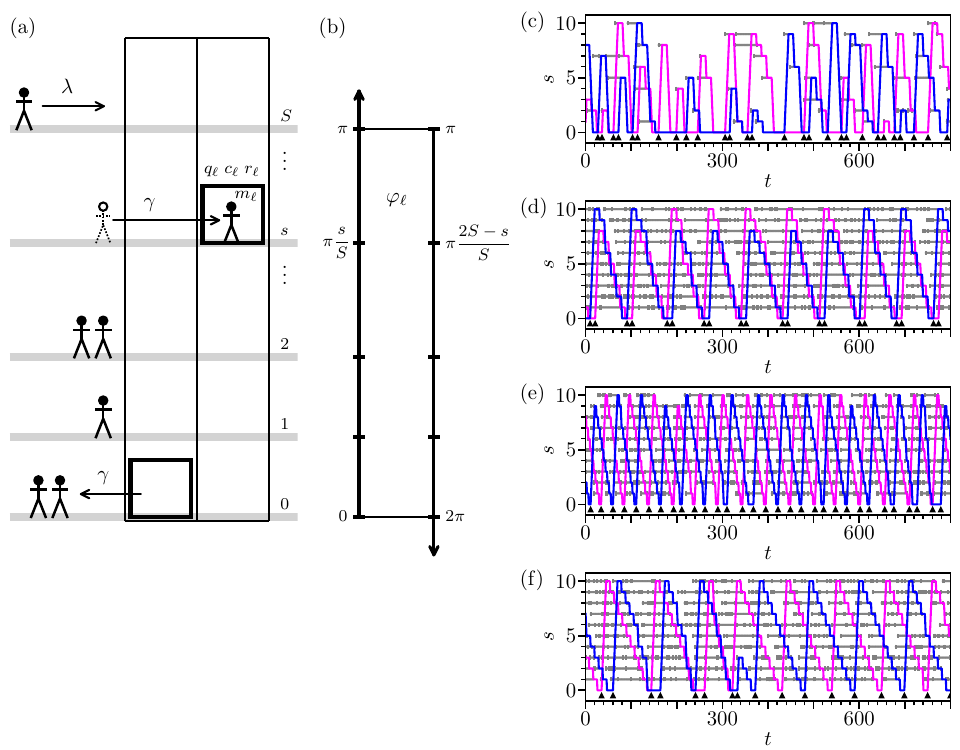}
	\caption{%
		Schematics of the model and resulting trajectories. 
		(a) The rule-based discrete model with two lifts and (b) the definition of phase $\varphi_{\ell}$. 
		(c--f)
		Trajectories of two lifts (blue and magenta lines) with $\gamma = 10$ for (c) $\mu=0.1$ and (d) $\mu=1$, (e) with $\gamma = 3$ for $\mu = 1$, and (f) with $\gamma=8$ for $\mu=1$. 
		On each floor the gray horizontal lines indicate the existence of waiting passengers and the gray vertical bars represents appearance of new passengers. 
		The black arrowheads indicate the departure times $\hat{t}_i$. 
	}%
	\label{fig:schematics}
\end{figure}
\subsection{Rule-based discrete model} \label{sec:model}

We start by briefly describing our rule-based discrete model (Fig~\ref{fig:schematics}a), the details of which are introduced  in the method section~\ref{sec:model_equations}. 
We consider a building that possesses $S +1$ floors, where the ground and highest floors are respectively given by $s=0$ and $s=S$. 
It is equipped with two lifts installed next to each other. 
For simplicity, we assume that all the passengers are heading to the ground floor, corresponding to the day-end of an office building or the closing time of a department store. 
We also assume the capacity of a lift is large enough so that all the waiting passengers can get on the lift that arrives. 
We discretize space by the floors and time by the time that a lift takes to move from one floor to the next. 
Then, at each time the position of lift takes an integer between 0 and $S$, which changes in time based on the following four basic rules. 
Firstly, once a lift starts to move upwards it does not stop nor move downwards until it reaches the highest floor among those calling it. 
Secondly, once a lift with finite riding passengers starts to move downwards it does not move upwards until it reaches the ground floor. 
Thirdly, a lift moves only when some passengers are on board or when it is called from other floors. 
Finally, when a lift comes to a floor with waiting passengers or to the ground floor, it stops there for a fixed time $\gamma$ to let the passengers get in or out. 

On each floor, passengers show up stochastically, the number of which is drawn from Poisson distribution
\begin{align}
	n_s^+ \sim P_\lambda(n) = \frac{\lambda^n}{n!} e^{-\lambda}
\end{align}
with $\lambda$ the influx rate of new passengers. 
We assume the influx rate of each floor is the same, $\lambda = \mu / S$, as is considered in the previous studies~\cite{Tanida2021Dynamic,Tanida2022The}. 
Here, $\mu$ is the influx rate of the entire building.
Since there are two lifts, the waiting passengers on each floor calls the one that is expected to arrive first. 

In this set up, the free parameters are the floor number $S$, stopping time $\gamma$, and the passenger influx rate $\mu$. 
Hereafter, we fix $S=10$, and vary $\gamma$ and $\mu$. 
For $\gamma = 10$, the two lifts move independently as in Fig.~\ref{fig:schematics}c for a small $\mu$, while for a large $\mu$ they move close to each other and synchronize their motion as shown in Fig.~\ref{fig:schematics}d.
That is, the dynamics of the two lifts transitions from disordered to synchronized states as the influx rate $\mu$ increases, which is consistent with the previous report~\cite{Tanida2021Dynamic}. 

\subsection{Effective interaction between lifts} \label{sec:interaction}
\begin{figure*}[tb]
	\centering
	\includegraphics[width=\textwidth]{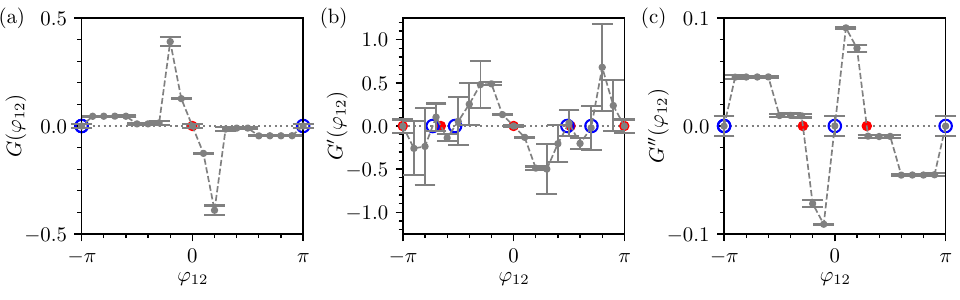}
	\caption{
		Effective interaction of lifts for $\mu = 1$ and $\gamma=10$. 
		(a) The measured coupling function $G(\varphi_{12})$, as well as (b) that for co-descent period $G^\prime(\varphi_{12})$ and (c) that for non-co-descent period $G^{\prime\prime}(\varphi_{12})$ are displayed against phase difference $\varphi_{12}$. 
		The red filled dots and blue open circles represent stable and unstable fixed points, respectively. 
		The stable fixed point of $G(\varphi_{12})$ and $G^\prime(\varphi_{12})$ at $\varphi_{12} = 0$ demonstrate the existence of  effective attractive interaction overall that originates from the co-descent period, while the unstable fixed point $\varphi_{12} = 0$ of $G^{\prime\prime}(\varphi_{12})$ indicate the existence of effective repulsive interaction during non-co-descent period. 
		The errorbars are for the ensemble average over 1000 data sets. 
	}
	\label{fig:interaction}
\end{figure*}
In order to investigate the mechanism behind this synchronization, we analyse quantitatively the effective interaction acting between the lifts. 
To this end, we define the phase of lift $\ell$ from its position $q_\ell(t)$ by 
\begin{align}
	\varphi_\ell = 2\pi \delta_{c_\ell, -1} +\pi \frac{q_\ell(t)}{S} c_\ell
	\label{eq:phi_ell}
\end{align}
Here $c_{\ell}$ is the state of lift $\ell$ that takes $c_{\ell} = 1$ for the ascent state, $c_{\ell} = -1$ for the descent state, and $c_{\ell} = 0$ for the rest state.
This phase $\varphi_\ell$ distinguishes not only the lift position but also if the lift ascends or descends. 
That is, it increases from zero to $\pi$ as a lift ascends from the ground floor to the highest floor $S$, and from $\pi$ to $2\pi$ as it descends from the floor $S$ to the ground floor. 
See Fig.~\ref{fig:schematics}b. 
Note that $\varphi_\ell$ is $2\pi$ periodic. 
By assuming a simple phase equation for $\varphi_{\ell}$ (eqs.~\eqref{eq:phase_eq1} and \eqref{eq:phase_eq2} in the method), we obtain the  equation for the phase difference $\varphi_{12} = \varphi_1 -\varphi_2$ as
\begin{align}
	\frac{d \varphi_{12}}{dt} = \omega(\varphi_1) -\omega(\varphi_2) + 2 G( \varphi_{12} )
	\label{eq:phase_diff_eq_varphi12}
\end{align}
where $\omega(\varphi_i)$ is the characteristic frequency that depends on  $\varphi_\ell$. 
From the time series of $\varphi_{1}$ and $\varphi_{2}$, we can estimate the coupling function $G( \varphi_{12} )$ by solving this equation for it (See the method for details). 

The coupling function $G( \varphi_{12} )$ that is measured from the time series data of lift motion generated by numerically solving the rule-based discrete model, is plotted in Fig.~\ref{fig:interaction}a for $\mu = 1$ and $\gamma = 10$. 
It crosses zero at $\varphi_{12} = 0$ with a negative slope, and thus, $\varphi_{12} = 0$ is a stable fixed point, which clearly demonstrates the existence of effective attractive interaction. 
Since all the passengers are heading towards the ground floor, the ascending and descending motions are not symmetric. 
Therefore, we separate $G(\varphi_{12})$ into $G^{\prime}( \varphi_{12} )$ and $G^{\prime\prime}( \varphi_{12})$;
$G^{\prime}( \varphi_{12} )$ is the coupling function of the co-descent period where both lifts are moving downwards ($\pi \le \varphi_1 < 2\pi$ and $\pi \le \varphi_2 < 2\pi$), whereas $G^{\prime\prime}( \varphi_{12})$ is for the non-co-descent period, which are displayed in Figs.~\ref{fig:interaction}b and c, respectively. 
$\varphi_{12} = 0$ is also a stable fixed point of $G^{\prime}( \varphi_{12} )$, which confirms that the effective attractive interaction originates from the co-descent period. 
Interestingly, however, the fixed point $\varphi_{12} = 0$ of $G^{\prime\prime}( \varphi_{12})$ is unstable, indicating that the effective repulsive interaction acts between the lifts during the non-co-descent period. 

These results suggest the following underlying mechanism. 
Both attractive and repulsive interactions effectively act between the two lifts. 
The attractive interaction occurs during the co-descent period ---to which the analogy to the bus-route model applies--- and the repulsive interaction acts during the non-co-descent period. 
When the attractive interaction dominates, the synchronization occurs as shown in Fig.~\ref{fig:schematics}d. 

If the above mechanism is true, one may wonder whether we can change the ratio of the effective attractive and repulsive interactions. 
In particular, can we make the situation in which the effective repulsive interaction dominates? 
Since the effective attractive and repulsive interactions act during the co-descent and non-co-descent period, this may be realized by changing the ratio of these periods. 
In fact, the average speed, i.e., the magnitude of the average slope of the trajectory in Fig.~\ref{fig:schematics}d, is smaller for descending than that for ascending since a descending lift stops at the floor with waiting passengers for $\gamma$. 
Therefore, one possibility is to decrease the stopping time $\gamma$, which reduces the difference in the average speed. 

The result is depicted in Fig.~\ref{fig:schematics}e for $\gamma = 3$, where the two lifts tend to depart the ground floor one after the other almost at the same intervals, indicating that the effective repulsive interaction dominates. 
Actually, the two lifts  tend to stay separated during the entire period, exhibiting anti-phase synchronization, as we identify quantitatively later in the next section. 
To distinguish the previous synchronization from this anti-phase synchronization, we refer to the one like in Fig.~\ref{fig:schematics}d as in-phase synchronization hereafter. 

\subsection{Transition between in-phase and anti-phase synchronization} \label{sec:control}
\begin{figure*}[tb]
	\centering
	\includegraphics[width=\textwidth]{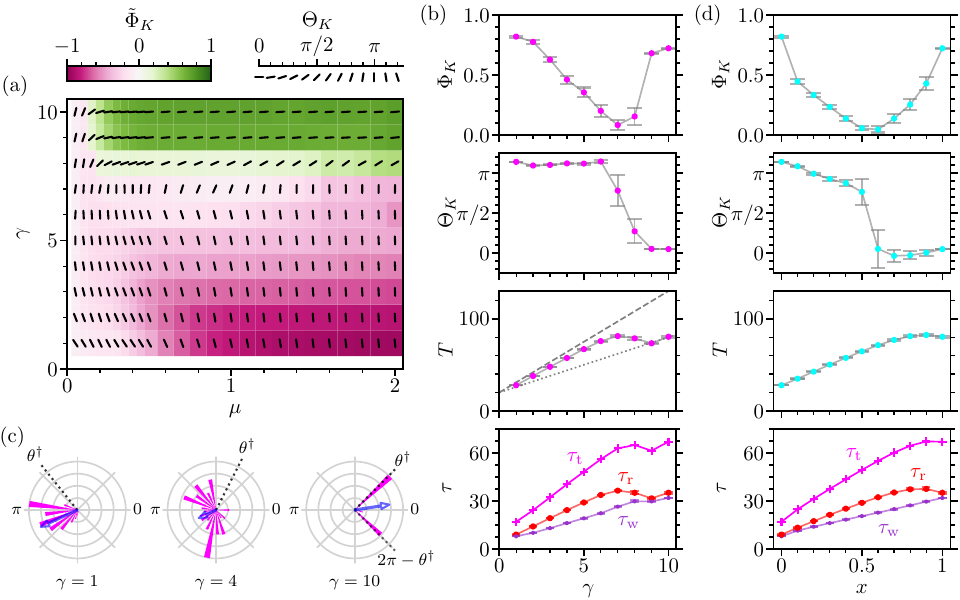}
	\caption{
		Transition of in-phase and anti-phase synchronization states. 
		(a) Phase diagram. 
		The colour represents the value of $\tilde{\Phi}_K = \Phi_K \,{\rm sign}{(\cos{\Theta_K})}$, and the black bars indicates the value of $\Theta_K$. 
		The thick green and magenta region correspond to in-phase and anti-phase synchronization states, respectively. 
		(b,d) Dependence of the synchronization state on (b) the stopping time $\gamma$ and (d) the rate $x$ of the stochastic stopping time. 
		Order parameters $\Phi_K$ and $\Theta_K$, round trip time $T$, and the mean times for waiting $\tau_{\rm w}$, riding $\tau_{\rm r}$, and travelling $\tau_{\rm t}$ are displayed. 
		The stochastic stopping time takes a value unity with the rate $1-x$ and 10 with the rate $x$. 
		Ensemble average is calculated over 100 samples and the error bars represent the standard deviation. 
		In the third panels for $T$, the dashed and dotted gray lines indicate the round trip time when a lift stops at all floors ($T^S$) and half of them ($T^{S/2}$), respectively. 
		(c) Distribution of the departure time intervals $\theta = 2\pi \mathit{\Delta}\hat{t} / T$ for $\mu=1$ and $\gamma = 1$ (left), 4 (middle), and 10 (right). 
		The purple arrow shows its average, i.e., $\Phi_K \exp{\left[ i \Theta_K \right]}$. 
		The black dotted line represents $\theta^\dag = 2\pi S/T$. 
		In the right panel for $\gamma=10$, its complement $2\pi -\theta^\dag$ is also plotted. 
	}
	\label{fig:gamma}
\end{figure*}
Now in order to characterize the dynamics of two lifts quantitatively and distinguish the in-phase and anti-phase synchronizations, we introduce the order parameters $\Phi_K$ and $\Theta_K$ which are the magnitude and angle of Kuramoto-type order parameter defined by the departure time intervals from the ground floor $\mathit{\Delta}\hat{t}$ normalized by the average round trip time $T$ (See the method for the detailed definition). 
That is, $\Phi_K$ and $\Theta_K$ measure how regularly the lifts depart from the ground floor and the average interval of departure times with respect to the round trip time, respectively. 

In Fig.~\ref{fig:gamma}a, we show phase diagram against the passenger influx rate $\mu$ and stopping time $\gamma$, where the colour and the black bars correspond to the value of $\tilde{\Phi}_K = \Phi_K \,\cos{\Theta_K} / \left|\cos{\Theta_K}\right|$ and $\Theta_K$, respectively. 
For a small $\mu$, the order parameter $\Phi_K$ takes significantly low value, where the two lifts depart from the ground floor randomly, as in Fig.~\ref{fig:schematics}c. 
In the region of the thick green for a large $\mu \gtrsim 0.3$ and $\gamma \ge 9$, $\Phi_K$ becomes large while $\Theta_K \sim 0$, which means that the two lifts departs from the ground floor regularly and almost simultaneously, resulting in in-phase synchronization as shown in Fig.~\ref{fig:schematics}d. 
On the other hand, in the thick magenta region, $\Phi_K$ is large and $\Theta_K$ is close to $\pi$, indicating that, although the two lifts depart from the ground floor also regularly, they are separated by almost half the round trip time ($\Theta_K \sim \pi$), 
resulting in anti-phase synchronization as depict in Fig.~\ref{fig:schematics}e. 

To proceed the analysis on the transition between in-phase and anti-phase synchronization, we plot the order parameters $\Phi_K$ and $\Theta_K$ as a function of $\gamma$ for $\mu=1$ in Fig.~\ref{fig:gamma}b. 
The value $\Phi_K$ is large for a large $\gamma$ ($\ge9$), while it suddenly drops close to zero at $\gamma =8$. 
By further decreasing $\gamma$, the order parameter $\Phi_K$ gradually increases again. 
On the other hand, the angle $\Theta_K$ changes from $0$ (for $\gamma > \gamma_c$) to $\pi$ (for $\gamma < \gamma_c$) more sharply at around $\gamma_c \approx 7$, where the error becomes large. 
These results indicate that the dynamics transitions at around $\gamma_c$ from in-phase synchronization for a larger $\gamma$, where the attractive interaction dominates, to anti-phase synchronization for a smaller $\gamma$, where the repulsive interaction dominates. 
Interestingly, around the transition point, the lifts exhibits bistability and in-phase and anti-phase synchronization coexist as shown in Fig.~\ref{fig:schematics}f. 

One may wonder why $\Phi_K$ changes gradually for smaller $\gamma$ around the threshold while the change in $\Theta_K$ is sharp. 
The reason can be understood from the distribution of the departure time intervals $\theta_i = 2\pi \mathit{\Delta}\hat{t}_i / T$, which is displayed in Fig.~\ref{fig:gamma}c. 
On the one hand, when in-phase synchronization occurs ($\gamma = 10$), $\theta_i$ is distributed at $\theta^\dag = 2\pi S / T$ and its complement $2\pi - \theta^\dag$. 
Note that $\theta^\dag$ corresponds to the time required for a lift to move from the ground floor to the $S$th floor. 
This means, when two lifts are at the ground floor and one of them starts to ascend, the other one is waiting to ascend until the former reaches the highest floor (Fig.~\ref{fig:schematics}d). 
On the other hand, when anti-phase synchronization occurs for $\gamma = 1$, $\theta_i$ is distributed around the average $\sim \pi$, instead of $\theta^\dag$. 
The distribution of the departure time intervals $\theta_i$ for the anti-phase synchronization is less sharp than that for the in-phase synchronization. 
This is probably because the effective repulsive interaction is local and only tend to keep the phase difference away from $\varphi_{12} = 0$ (Fig~\ref{fig:interaction}c). 
The distribution of $\theta_i$ for the anti-phase synchronization is also rather noisy because the lift motion is induced by stochastic passenger influx that is strongly fluctuating. 
As $\gamma$ increases, the distribution becomes much broader (see the middle panel of Fig.~\ref{fig:gamma}c for $\gamma=4$), resulting in the decrease of its average value $\Phi_K$. 
Again, this can be understood as the decrease of the impact of the effective repulsive interaction. 
Note that $\theta_i$ is still not distributed at around $\theta^\dag$. 
From these we conclude that the increase in width of the distribution of departure time interval $\theta_i$ is the reason why the order parameter $\Phi_K$ decreases only gradually with $\gamma$ before the transition. 

The transition between the anti-phase to in-phase synchronization also affects the round trip time $T$ as shown in Fig.~\ref{fig:gamma}b, which shifts from $T^S = 2 S +\gamma (1 +S)$ (the gray dashed line) to $T^{S/2} = 2 S +\gamma ( 1 + S/2 )$ (the gray dotted line) around $\gamma_c$. 
Here, $T^S$ is the time required for a lift stopping at all the floors while descending, whereas $T^{S/2}$ is the time required for a lift stopping at only half the floors, that happens for the in-phase synchronization state where each lift stops at every two floors because it can pass the floor where the other one is serving (See Fig.~\ref{fig:schematics}d). 
As shown in Fig.~\ref{fig:gamma}b, this shift in the round trip time has less impact on the average waiting time $\tau_{\rm w}$, i.e. the time required for passengers to spend on each floor from when they show up to when they get on a lift. 
On the other hand, it affects the average riding time $\tau_{\rm r}$, and thus, the travelling time $\tau_{\rm t}$ given by the sum of the waiting and riding times $\tau_{\rm t} = \tau_{\rm w} +\tau_{\rm r}$. 

So far, we confined ourselves to the case of constant stopping time. 
Finally, as a more realistic situation, we consider the case where the stopping time $\gamma$ varies.
In order to keep the situation still simple, however, we consider the case in which the stopping time $\gamma$ is determined stochastically and takes the value unity with the rate $1-x$ and 10 with the rate $x$. 
This corresponds to the situation where riding passengers sometimes press the close button. 
The results are summarized in Fig.~\ref{fig:gamma}d for $\mu = 1$. 
By increasing $x$, the order parameter $\Theta_K$ transitions from $\pi$ to 0 around $x \sim 0.6$, where $\Phi_K$ is almost vanishing. 
Around this transition, however, the round trip time $T$ increases smoothly, indicating that the in-phase and anti-phase synchronization transition is not caused by the discrete change in $T$. 

\section{Discussion and conclusion} \label{sec:summary}
To summarize, we have investigated the mechanism behind the synchronization of two lifts by analyzing the time series that are numerically generated based on a rule-based discrete model. 
By measuring the effective interaction acting between the lifts, we have revealed the existence of both attractive and repulsive interactions and the synchronization originates from the competition between them. 
Moreover, by varying the parameter of the microscopic rule-based model, we were successful to tune the impact of these competing interactions, and realize the transition from in-phase to anti-phase synchronization. 
We have also showed that the transition occurs in a realistic case where the stopping time is determined stochastically. 

The attractive interaction acts during the co-descent period where both lifts move downwards, and while doing so they stop at the floor where passengers are waiting. 
Thus, its origin can be understood in the analogy to the bus route model. 
On the other hand, the repulsive interaction act during the non-co-descent period.
Therefore, we presume that it originates from the facts that without riding passengers a lift moves only when it is called and that each floor calls the lift that is estimated to arrive first. 
More precisely, suppose that both lifts are at the ground floor without riding passengers. 
If one of them is called and starts to ascend, the other one is less likely to be called for the next several time since the first one is expected to arrive faster to any floors. 
This tends to keep the second lift at the ground floor while the first one ascends, resulting in the effective repulsive interaction. 

Since the lift system considered in this study is simple and idealized, further complexity can be included to make it more realistic. 
For instance, the stopping time can also depend on the number of passengers getting in or out, although we set it as a constant or to be chosen stochastically from two constant values. 
In addition, in real situation, the destination of passengers may vary and some want to get off at an intermediate floor or even want to go up. 
In fact, some previous studies \cite{Nagatani2003Complex, Feng2021When} suggest lifts also exhibit synchronization at the start-of-day situation, when all the passengers are travelling from the ground to higher floors. 

Finally, it is of practical interest if the synchronization improves the efficiency of transportation. 
Although it depends on how the efficiency is defined, our results show that for a large stopping time the round trip time, and thus the average travelling time, becomes smaller when in-phase synchronization occurs compared to the values expected if the anti-phase synchronization continues (See Fig.~\ref{fig:gamma}b). 
Although further analysis on the efficiency is required and it is beyond the scope of this study, we believe our current results, in particular the existence of both attractive and repulsive interactions, provide a crucial knowledge for such an application aspect because it is of great importance to know the actual mechanism at play to design efficient transportation system. 

In conclusion, our study revealed the emergence of unexpected synchronization dynamics of two-lift system, which transitions between in-phase and anti-phase synchronization states, and the underlying mechanism is understood as a competition of the effective attractive and repulsive interactions. 
Since lifts are man-made system, one may presume their dynamics is fully controlled by the system design. 
However, the results of our study indicate that the lifts are coupled through waiting passengers and show a feature of non-equilibrium systems and emergent dynamics. 
Lift move back-and-forth along its track periodically and its oscillatory dynamics depends on passengers appearing stochastically. 
This resembles a molecular motor such as F1-ATPase molecular motor~\cite{Noji1997Direct}, which possesses a specific path of molecular conformation change and which is driven along the path by stochastic interaction with adenosine triphosphate (ATP). 
In this sense, the lift dynamics is a simple example of active matter in social systems. 
Therefore, we expect that our study opens a new avenue to social active matter. 

\section{Methods}\label{sec:methods}
\subsection{Rule-based discrete model}\label{sec:model_equations}
Here we describe the details of our rule-based discrete model (See Fig~\ref{fig:schematics}a). 
In order to realize the four rules for the lift motion explained in the main text, we first introduce the state $c_{\ell}(t)$ of lift $\ell$, which takes either of the three states; 
the ascent state ($c_{\ell}(t) = 1$),
the descent state ($c_{\ell}(t) = -1$), and
the rest state ($c_{\ell}(t) = 0$). 
The state $c_{\ell}(t)$ of lift $\ell$ is determined as follows 
by using the number of riding passengers $m_\ell(t)$ and the quantity $\chi_{\ell}^{s}(t)$ that takes a value unity if it is called from floor $s$ and zero otherwise; 
$c_{\ell}(t) = 1$ if $c_{\ell}(t-1) \ge 0$ and $m_{\ell}(t) = 0 $ and $\sum_{s > q_{\ell}(t-1)} \chi_{\ell}^{s}(t) > 0$; 
$c_{\ell}(t) = -1$ if $\sum_{s > q_{\ell}(t-1)} \chi_{\ell}^{s}(t) = 0$ and $\sum_{s < q_{\ell}(t-1)} \chi_{\ell}^{s}(t) > 0$ or if $m_{\ell}(t) > 0$ ; 
$c_{\ell}(t) = 0$ otherwise. 
The position $q_\ell(t)$ of lift $\ell$ is updated as
\begin{align}
	q_\ell(t) = q_\ell (t-1) +c_\ell(t) \delta_{ r_\ell(t), 0 }
	\label{eq:dq_ell}
\end{align}
where $\delta_{a,b}$ is Kronecker's delta that is 1 if $a=b$ and 0 otherwise. 
$r_\ell(t)$ is the remaining stopping time at the current floor, that counts down if $r_\ell(t)>0$ as 
\begin{align}
	r_\ell(t +1) = r_\ell(t) -1
	\label{eq:r_ell}
\end{align}
If $r_\ell(t)=0$, however, it is reset to a constant $\gamma>0$ only when lift $\ell$ arrives at a new floor with waiting passengers to let them in or at the ground floor to let them out. 
More precisely, this resetting occurs when both $r_\ell(t)=0$ and one of the following four conditions is satisfied; 
(a) $c_\ell(t) = 1$ and $q_{\ell}(t) = \hat{s}_{\ell}(t)$, 
(b) $c_\ell(t) = -1$ and $n_{q_{\ell}(t)}(t) > 0$, 
(c) $c_\ell(t) = -1$ and $q_{\ell}(t) = 0$, or 
(d) $c_{\ell}(t) = 0$ and $n_{q_{\ell}(t)}(t) > 0$. 
Here $\hat{s}_{\ell}(t) = \argmax{s}{ \left[ s \chi_{\ell}^s(t) \right] }$ is the highest floor calling lift $\ell$, where the function $\argmax{x}{\left[ y(x) \right]}$ gives $x$ that maximizes $y(x)$.  

Since there are two lifts, each floor with a finite waiting passengers calls the lift that is expected to arrive first. 
By defining
\begin{align}
	&
	\check{\tau}( s_1, s_2, t) 
	= 
	|s_1 -s_2| +\sum_{s^\prime = s_2+1}^{s_1-1} \gamma H(n_{s^{\prime}}(t)) H(s_1 -s_2 -1)
	\label{eq:check-tau_ell^s(t)}
\end{align}
using the Heaviside step function $H(x)$ that takes 1 if $x>0$ and otherwise 0, 
the estimated arrival time $\tau_{\ell}^s(t)$ of lift $\ell$ at floor $s$ is calculated as 
\begin{align}
	&
	\tau_{\ell}^s(t) = 
	\notag\\&
	\left\{
	\begin{array}{cl}
		0
		& \text{if~} c_{\ell}(t) \le 0 \text{~and~} q_{\ell}(t) = s \\
		\displaystyle
		r_{\ell}(t) +\check{\tau}( q_{\ell}(t), s, t)
		& \text{if~} c_{\ell}(t) \le 0 \text{~and~} q_{\ell}(t) > s \\
		r_{\ell}(t) 
		+\check{\tau}( q_{\ell}(t), s^{\dag}(t), t)
		+\check{\tau}( s^{\dag}(t), s, t)
		+\gamma
		& \text{if~} c_{\ell}(t) = 0 \text{~and~} q_{\ell}(t) < s 
		\\ &
		\text{~or if~} c_{\ell}(t) > 0 \\
		r_{\ell}(t) 
		+\check{\tau}( q_{\ell}(t), 0, t)
		+\check{\tau}( 0, s^{\dag}(t), t)
		+\check{\tau}( s^{\dag}(t), s, t)
		+2\gamma
		& \text{if~} c_{\ell}(t) = -1 \text{~and~} q_{\ell}(t) < s 
	\end{array}
	\right.
	\label{eq:tau_ell^s(t)}
\end{align}
where
$s^{\dag}(t)$ represents the highest floor on which at least one passenger is waiting. 
If no passenger is waiting at floor $s$, we set $\tau_{\ell}^s(t) = \infty$ for all $\ell$. 
Then, if $\tau_{\ell}^s(t) < \tau_{\ell^\prime}^s(t)$,  we set $\chi_{\ell}^s(t) = 1$ and $\chi_{\ell^\prime}^s(t) = 0$. 
In the case of the equivalent estimated arrival time $\tau_{\ell}^s(t) = \tau_{\ell^\prime}^s(t)$, if there is a lift called at the previous time, the same one is still called, i.e., $\chi_{\ell}^s(t) = \chi_{\ell}^s(t-1)$, but if neither was called ($\chi_{0}^s(t-1) = \chi_{1}^s(t-1) = 0$) then one is chosen randomly. 
If there are no waiting passenger on floor $s$, no lift is called; 
$\chi_{0}^s(t) = \chi_{1}^s(t) = 0$. 
In the case of stochastic stopping time, the estimated arrival time $\tau_\ell^s(t)$ is calculated for $\gamma = 10$ regardless of the value of $x$.

\subsection{Phase equations}\label{phase_equations}
For the phase $\varphi_{\ell}$ defined by eq.~\eqref{eq:phi_ell}, we assume that its time evolution is governed by phase equations 
\begin{gather}
	\frac{d \varphi_1}{dt} = \omega(\varphi_1) + G( \varphi_1 -\varphi_2 )
	,~
	\label{eq:phase_eq1}
	\\
	\frac{d \varphi_2}{dt} = \omega(\varphi_2) - G( \varphi_1 -\varphi_2 )
	.
	\label{eq:phase_eq2}
\end{gather}
Here, since the speed of the lift is different between the ascent and descent periods, the characteristic frequency $\omega(\varphi_i)$ depends on the value of $\varphi_i$; 
$\omega(\varphi_i) = \pi/S$ for $0 \le \varphi_i < \pi$, whereas for $\pi \le \varphi_i < 2\pi$ it is set as 
$\omega(\varphi_i) = \pi / S (1+\gamma)$.
The latter is rigorously true for a high influx rate $\mu$, in which case the lift stops at all floors. 
Then, the phase difference $\varphi_{12} = \varphi_1 -\varphi_2$ obeys
\begin{align}
	\frac{d \varphi_{12}}{dt} = \omega(\varphi_1) -\omega(\varphi_2) + 2 G( \varphi_{12} )
	\label{eq:phase_diff_eq}
\end{align}
Therefore, the coupling function $G( \varphi_{12})$ can be estimated from the time series of $\varphi_1$ and $\varphi_2$ by
\begin{align}
	G( \varphi_{12} ) 
	&
	= \left\langle \frac{1}{2} \left[  \varphi_{12}(t +1) -\varphi_{12}(t)  - \left\{ \omega(\varphi_1(t)) -\omega(\varphi_2(t)) \right\} \right] \right\rangle
	\label{eq:coupling_function}
\end{align}
where the average $\langle \cdot \rangle$ is calculated over time. 
Although $\varphi_\ell$ defined by eq.~\eqref{eq:phi_ell} does not always reach $\pi$ since a lift ascending due to the demand may start descending before reaching the floor $S$, we expect this has only a minor effect for a large $\mu$ in particular for the phase difference. 

For each time series, we also measure the coupling function $G^{\prime}( \varphi_{12} )$ for the co-descent period where both lifts are moving downwards ($\pi \le \varphi_1 < 2\pi$ and $\pi \le \varphi_2 < 2\pi$) and $G^{\prime\prime}( \varphi_{12})$ for the non-co-descent period. 
To measure $G^{\prime}( \varphi_{12} )$ and $G^{\prime\prime}( \varphi_{12} )$, we first split the time series into the co-descent and non-co-descent periods based on the values of $\varphi_1(t)$ and $\varphi_2(t)$, and calculate the coupling function using the time-series data of each period using eq.~\eqref{eq:coupling_function}. 

\subsection{Kuramoto-type order parameter}\label{order-parameter}
We define the order parameters $\Phi_K$ and $\Theta_K$ as the magnitude and angle of Kuramoto-type order parameter defined by
\begin{align}
	\Phi_K \exp{[ i \Theta_K ]} = 
	\left\langle \exp{\left[2\pi i \mathit{\Delta}\hat{t}_j /T\right]} \right\rangle_j
	\label{eq:Kuramoto_op}
\end{align}
Here, the departure time interval $\mathit{\Delta}\hat{t}_j = \hat{t}_{j+1} -\hat{t}_j$ is calculated using a series of departure time of any lift from the ground floor $\{ \hat{t}_{j} \}$ ($ \hat{t}_{j} \le \hat{t}_{j+1}$) with $j$ representing the chronological order, over which the average $\langle \cdot \rangle_j$ is calculated. 
The round trip time $T$ is defined by
$
T = \left\langle \check{t}^\ell_j +\gamma -\hat{t}^\ell_j \right\rangle_{j,\ell}
$
where 
$\hat{t}^\ell_j$ ($\check{t}^\ell_j$) is the departure (arrival) time of $j$th round trip from (to) the ground floor, and
the average $\langle \cdot \rangle_{j,\ell}$ is calculated over $j$ and $\ell$. 

\begin{acknowledgments}
We thank H. Kori for helpful discussions. 
JSPS Core-to-Core Program ``Advanced core-to-core network for the physics of self-organizing active matter (JPJSCCA20230002)'' is acknowledged. 

MT is supported by JSPS KAKENHI Grant Numbers JP22K14017 and JP24H01485. 
ST is supported by JSPS KAKENHI Grant Number JP25K17790, the Odakyu Foundation Research Program, the Obayashi Foundation Research Program, and JST PRESTO Grant Number JPMJPR25KE.

The authors declare no competing interests.

[Author contribution] 
MT: Design of the work, modelling, analysis, interpretation of data, drafting the article. 
ST: Design of the work, interpretation of data. 

\end{acknowledgments}


\begin{thebibliography}{10}


\bibitem{Vicsek2012Collective}
T.~Vicsek, A.~Zafeiris, Collective motion.
\newblock Physics Reports \textbf{517}(3 - 4), 71 -- 140 (2012).

\bibitem{Elgeti2015Physics}
J.~Elgeti, R.G. Winkler, G.~Gompper, Physics of microswimmers?single particle
motion and collective behavior: a review.
\newblock Reports on Progress in Physics \textbf{78}(5), 056601 (2015).

\bibitem{teVrugt2025Metareview}
M.~Te~Vrugt, R.~Wittkowski, Metareview: a survey of active matter reviews.
\newblock The European Physical Journal E \textbf{48}(3), 12 (2025)

\bibitem{Liebchen2022Interactions}
B.~Liebchen, A.K. Mukhopadhyay, Interactions in active colloids.
\newblock Journal of Physics: Condensed Matter \textbf{34}(8), 083002 (2021).

\bibitem{Sanchez2012Spontaneous}
T.~Sanchez, D.T. Chen, S.J. DeCamp, M.~Heymann, Z.~Dogic, Spontaneous motion in
hierarchically assembled active matter.
\newblock Nature \textbf{491}(7424), 431--434 (2012)

\bibitem{Sekine2024Emergence}
S.~Sekine, M.~Tarama, H.~Wada, M.M. Sami, T.~Shibata, S.~Hayashi, Emergence of
periodic circumferential actin cables from the anisotropic fusion of actin
nanoclusters during tubulogenesis.
\newblock Nature Communications \textbf{15}(1), 464 (2024)

\bibitem{Tarama2022Mechanochemical}
M.~Tarama, K.~Mori, R.~Yamamoto, Mechanochemical subcellular-element model of
crawling cells.
\newblock Frontiers in Cell and Developmental Biology \textbf{10} (2022).

\bibitem{Bechinger2016Active}
C.~Bechinger, R.~Di~Leonardo, H.~L\"owen, C.~Reichhardt, G.~Volpe, G.~Volpe,
Active particles in complex and crowded environments.
\newblock Rev. Mod. Phys. \textbf{88}, 045006 (2016).

\bibitem{Nakata2018Self-organized}
S.~Nakata, V.~Pimienta, I.~Lagzi, H.~Kitahata, N.J. Suematsu,
\emph{Self-organized {Motion}: {Physicochemical} {Design} based on
	{Nonlinear} {Dynamics}} (The Royal Society of Chemistry, 2018).

\bibitem{Gu2025Emergence}
F.~Gu, B.~Guiselin, N.~Bain, I.~Zuriguel, D.~Bartolo, Emergence of collective
oscillations in massive human crowds.
\newblock Nature \textbf{638}(8049), 112--119 (2025)

\bibitem{Doostmohammadi2023Editorial}
A.~Doostmohammadi, M.G. Mazza, T.N. Shendruk, H.~Stark, Editorial: Active and
intelligent living matter: from fundamentals to applications.
\newblock Frontiers in Physics \textbf{Volume 11 - 2023} (2023).

\bibitem{teVrugt2025Artificial}
M.~te~Vrugt, \emph{Artificial intelligence and intelligent matter} (Springer,
2025)

\bibitem{Dreyer2025Comparing}
T.~Dreyer, A.~Haluts, A.~Korman, N.~Gov, E.~Fonio, O.~Feinerman, Comparing
cooperative geometric puzzle solving in ants versus humans.
\newblock Proceedings of the National Academy of Sciences \textbf{122}(1),
e2414274121 (2025).

\bibitem{Hakim2017Collective}
V.~Hakim, P.~Silberzan, Collective cell migration: a physics perspective.
\newblock Reports on Progress in Physics \textbf{80}(7), 076601 (2017).

\bibitem{Winfree1967Biological}
A.T. Winfree, Biological rhythms and the behavior of populations of coupled
oscillators.
\newblock Journal of Theoretical Biology \textbf{16}(1), 15--42 (1967).

\bibitem{Kuramoto1975Self-entrainment}
Y.~Kuramoto, \emph{Self-entrainment of a population of coupled non-linear
	oscillators}, in \emph{International Symposium on Mathematical Problems in
	Theoretical Physics}, ed. by H.~Araki (Springer Berlin Heidelberg, Berlin,
Heidelberg, 1975), pp. 420--422

\bibitem{Sakaguchi1986A}
H.~Sakaguchi, Y.~Kuramoto, A soluble active rotater model showing phase
transitions via mutual entertainment.
\newblock Progress of Theoretical Physics \textbf{76}(3), 576--581 (1986).

\bibitem{Reichert2005Synchronization}
M.~Reichert, H.~Stark, Synchronization of rotating helices by hydrodynamic
interactions.
\newblock The European Physical Journal E \textbf{17}(4), 493--500 (2005)

\bibitem{Uchida2010Synchronization}
N.~Uchida, R.~Golestanian, Synchronization and collective dynamics in a carpet
of microfluidic rotors.
\newblock Phys. Rev. Lett. \textbf{104}, 178103 (2010).

\bibitem{Levis2019Activity}
D.~Levis, I.~Pagonabarraga, B.~Liebchen, Activity induced synchronization:
Mutual flocking and chiral self-sorting.
\newblock Phys. Rev. Res. \textbf{1}, 023026 (2019).

\bibitem{Scholz2021Surfactants}
C.~Scholz, A.~Ldov, T.~P\"oschel, M.~Engel, H.~L\"owen, Surfactants and
rotelles in active chiral fluids.
\newblock Science Advances \textbf{7}(16), eabf8998 (2021).

\bibitem{Vahabli2023Emergence}
D.~Vahabli, T.~Vicsek, Emergence of synchronised rotations in dense active
matter with disorder.
\newblock Communications Physics \textbf{6}(56) (2023).

\bibitem{Reinhart-King2008Cell-Cell}
C.A. Reinhart-King, M.~Dembo, D.A. Hammer, Cell-cell mechanical communication
through compliant substrates.
\newblock Biophysical Journal \textbf{95}(12), 6044--6051 (2008).

\bibitem{Nitsan2016Mechanical}
I.~Nitsan, S.~Drori, Y.E. Lewis, S.~Cohen, S.~Tzlil, Mechanical communication
in cardiac cell synchronized beating.
\newblock Nature Physics \textbf{12}(5), 472--477 (2016)

\bibitem{Sepulchre2005Cooperative}
R.~Sepulchre, D.~Paley, N.~Leonard, \emph{Collective Motion and Oscillator
	Synchronization} (Springer Berlin Heidelberg, Berlin, Heidelberg, 2005), pp.
189--205.

\bibitem{Riedl2023Synchronization}
M.~Riedl, I.~Mayer, J.~Merrin, M.~Sixt, B.~Hof, Synchronization in collectively
moving inanimate and living active matter.
\newblock Nature Communications \textbf{14}(1), 5633 (2023)

\bibitem{Eckhardt2007Modeling}
B.~Eckhardt, E.~Ott, S.H. Strogatz, D.M. Abrams, A.~McRobie, Modeling walker
synchronization on the millennium bridge.
\newblock Phys. Rev. E \textbf{75}, 021110 (2007).

\bibitem{Poschel1994Synchronization}
T.~P\"oschel, J.A.C. Gallas, Synchronization effects in the dynamical behavior
of elevators.
\newblock Phys. Rev. E \textbf{50}, 2654--2659 (1994).

\bibitem{Hikihara1997Emergent}
T.~Hikihara, S.~Ueshima, Emergent synchronization in multi-elevator system and
dispatching control.
\newblock IEICE TRANSACTIONS on Fundamentals \textbf{E80-A}(9), 1548--1533
(1997)

\bibitem{Tanida2021Dynamic}
S.~Tanida, Dynamic behavior of elevators under random inflow of passengers.
\newblock Phys. Rev. E \textbf{103}, 042305 (2021).

\bibitem{Tanida2022The}
S.~Tanida, The synchronization of elevators when not all passengers will ride
the first-arriving elevator.
\newblock Physica A: Statistical Mechanics and its Applications \textbf{605},
127957 (2022).

\bibitem{Feng2021When}
Z.~Feng, S.~Redner, When will an elevator arrive?
\newblock Journal of Statistical Mechanics: Theory and Experiment
\textbf{2021}(4), 043403 (2021).

\bibitem{Nagatani2003Complex}
T.~Nagatani, Complex behavior of elevators in peak traffic.
\newblock Physica A: Statistical Mechanics and its Applications
\textbf{326}(3), 556--566 (2003).

\bibitem{Helbing2001Traffic}
D.~Helbing, Traffic and related self-driven many-particle systems.
\newblock Rev. Mod. Phys. \textbf{73}, 1067--1141 (2001).

\bibitem{Chowdhury2000Statistical}
D.~Chowdhury, L.~Santen, A.~Schadschneider, Statistical physics of vehicular
traffic and some related systems.
\newblock Physics Reports \textbf{329}(4), 199--329 (2000).

\bibitem{OLoan1998Jamming}
O.J. O'Loan, M.R. Evans, M.E. Cates, Jamming transition in a homogeneous
one-dimensional system: The bus route model.
\newblock Phys. Rev. E \textbf{58}, 1404--1418 (1998).

\bibitem{Sugiyama2008Traffic}
Y.~Sugiyama, M.~Fukui, M.~Kikuchi, K.~Hasebe, A.~Nakayama, K.~Nishinari, S.i.
Tadaki, S.~Yukawa, Traffic jams without bottlenecks?experimental evidence
for the physical mechanism of the formation of a jam.
\newblock New Journal of Physics \textbf{10}(3), 033001 (2008).

\bibitem{Noji1997Direct}
H.~Noji, R.~Yasuda, M.~Yoshida, K.~Kinosita~Jr, Direct observation of the
rotation of f1-atpase.
\newblock Nature \textbf{386}(6622), 299--302 (1997)

\end{thebibliography}
\end{document}